\documentclass[a4paper, usenatbib]{mnras}
\usepackage{graphicx}
\usepackage{epsfig}
\usepackage{subfig}
\usepackage{multirow}
\usepackage{mathrsfs,amsmath} 
\usepackage[utf8]{inputenc}

\title[Calibration Requirements for EoR Observations - II]{Calibration requirements for Epoch of Reionization 21-cm signal observations - II. Analytical estimation of the bias and variance with time-correlated residual gains}

\author[J. Kumar et al.]{Jais Kumar$^{1}$\thanks{E-mail: jaisk.rs.phy16@iitbhu.ac.in},
Prasun Dutta$^{1}$\thanks{E-mail:pdutta.phy@iitbhu.ac.in},
Samir Choudhuri$^{2}$,
Nirupam Roy$^{3}$
\\
$^{1}$Department of Physics, Indian Institute of Technology (Banaras Hindu University), Varanasi - 221005, India\\
$^{2}$Astronomy
Unit, Queen Mary University of London, Mile End Road, London E1 4NS, United Kingdom \\
$^{3}$Department of Physics, Indian Institute of Science, Bangalore, 560012, India
}
\date{Accepted XXX. Received YYY; in original form ZZZ}
\pubyear{2020}

\begin{document}

\def\HI{H{\sc~i}~}

\label{firstpage}
\pagerange{\pageref{firstpage}--\pageref{lastpage}}
\maketitle

\begin{abstract}
Observation of redshifted 21-cm signals from neutral hydrogen holds the key to understanding the structure formation and its evolution during the reionization and post-reionization era. Apart from the presence of orders of magnitude larger foregrounds in the observed frequency range, the instrumental effects of the interferometers combined with the ionospheric effects present a considerable challenge in the extraction of 21-cm signals from strong foregrounds. The systematic effects of time and frequency correlated residual gain errors originating from the measurement process introduce a bias and enhance the variance of the power spectrum measurements. In this work, we study the effect of time-correlated residual gain errors in the presence of strong foreground. We present a method to produce analytic estimates of the bias and variance in the power spectrum. We use simulated observations to confirm the efficacy of this method and then use it to understand various effects of the gain errors. We find that as the standard deviation in the residual gain errors increases, the bias in the estimation supersedes the variance. It is observed that an optimal choice of the time over which the gain solutions are estimated minimizes the risk. We also find that the interferometers with higher baseline densities are preferred instruments for these studies. 

\end{abstract}

\section{Introduction}
The dynamics, evolution, and thermal state of the Universe through cosmic time can be understood by studying the distribution of neutral hydrogen content in it. A particularly interesting era in the evolution of the Universe is the era when the first luminous sources were formed, and the radiation from them ionized the Universe. The ionization started as the first luminous sources start to come into existence; we call this epoch the Cosmic Dawn (CD henceforth). As the ionization progressed, the ionized regions around the luminous objects grew and started merging with each other, resulting in a complete ionization of the Universe by the redshift z $\sim6$. The cosmic time that followed the CD during which the ionization process prevailed is known as the Epoch of Reionization (EoR henceforth). Constraints on the redshift range of the reionization come from various studies such as the measurement of Gunn-Peterson troughs of quasars \citep{2001AJ....122.2850B, 2006AJ....132..117F}, the optical depth for Thomson scattering from Cosmic Microwave Background (CMB) polarization anisotropy \citep{2013ApJS..208...19H, 2020A&A...641A...6P}, IGM temperature measurements \citep{2002ApJ...567L.103T, 2010MNRAS.406..612B} etc. As the reionization process depends on the properties of the first luminous objects, the physical condition of the IGM, the matter density distribution, an observational probe to the neutral hydrogen distribution, and its evolution from the EoR holds the key to understanding the physics of this cosmic era. The 21-cm line emission from the neutral hydrogen provides a useful tool to study the \HI at different redshifts, including that in the EoR \citep{1997ApJ...475..429M, 1999A&A...345..380S, 2006PhR...433..181F, 2012RPPh...75h6901P}. As the 21-cm signal from the EoR is rather weak, its power spectrum or, alternatively, intensity mapping of the 21-cm signal holds the key for a successful observation. Several experiments are aimed to detect the fluctuating component of the redshifted 21 cm emission from the EoR with low-frequency radio telescopes including Giant Metrewave Radio Telescope \citep{1991CuSc...60...95S}, Low-Frequency Array \citep[LOFAR;][]{2013A&A...556A...2V}, Murchison Widefield Array \citep[MWA;][]{2013PASA...30....7T, 2013PASA...30...31B}, the Donald C. Backer Precision Array for Probing the Epoch of Reionization \citep[PAPER;][]{2010AJ....139.1468P, 2015ApJ...809...61A}, the Hydrogen Epoch of Reionization Array \citep[HERA;][]{2017PASP..129d5001D}, the Square Kilometer Array \citep{2013ExA....36..235M, 2015aska.confE...1K}.
Multiple experiments are also working on detecting the sky-averaged (global) \HI signal from the EoR. These include Experiment to Detect the Global EoR Signature \citep[EDGES;][]{2018Natur.555...67B} and Shaped Antenna measurement of the background RAdio Spectrum \citep[SARAS;][]{2018ApJ...858...54S}. \citet{2018Natur.555...67B} has reported the first tentative detection of the 21-cm absorption from the CD at a redshift of $17$. Here we focus on experiments that aim to detect the power spectrum of the \HI 21-cm signal from the EoR.

The presence of four to five orders of magnitude stronger emission from the astrophysical foregrounds at the observation frequency of the redshifted 21-cm emission is one of the major challenges in its detection \citep{1999A&A...345..380S, 2008MNRAS.385.2166A}. These foreground emissions include the radiation from the compact sources such as the radio galaxies as well as the diffuse synchrotron and free-free emissions from the Galaxy \citep{2005ApJ...625..575S, 2008MNRAS.385.2166A, 2012MNRAS.426.3295G, 2016JApA...37...35A, 2020MNRAS.494.1936C}. Different techniques are discussed in the literature to mitigate the effect of foreground: foreground avoidance \citep{2010ApJ...724..526D}, foreground mitigation \citep{2017NewA...57...94C}, foreground suppression \citep{2016MNRAS.463.4093C, 2019MNRAS.483.3910C, 2019MNRAS.483.5694B} are a few. In practice various foreground removal algorithms are investigated with the LOFAR-EoR data (FastICA \footnote {Fast Independent Component Analysis}, GMCA \footnote{Generalized Morphological Component Analysis} and GPR \footnote{Gaussian Process Regression}) in \citet{2021MNRAS.500.2264H}.

In an interferometric observation, the recorded data is modified by various instrumental effects. The calibration process defines the instrument's response to the incident signal parametrized as the per antenna complex gains. In general, the instrumental effects (which partially includes ionospheric effects) depend both on time and frequency. In this work, we are not considering any direction-dependent and baseline-dependent effects.
The calibration methods include observation of the standard calibrator sources, self-calibration, redundancy calibration, direction-dependent calibrations, etc. \citep{1984ARA&A..22...97P, 1992ExA.....2..203W,  2007ITSP...55.4497V, 2009ITSP...57.3512W}. In actual observations, the estimated gain solutions are not exact due to the rapidly fluctuating ionosphere, thermal noise in the visibilities, uncertainties in the sky models, and instrumental effects. They are always left with some residual gain errors.

Several sources of calibration/gain errors that lead to restrictions in the detection of the redshifted 21-cm signal are investigated in the literature.
\citet{2012ApJ...752..137M} have demonstrated that the simple frequency-independent calibration errors lead to residual power spectrum shapes contaminating nearly all k modes.
\citet{2016MNRAS.463.4317P} studied the systematic effects arising due to calibration and subtraction of bright point sources in the LOFAR-EoR residual data. These effects include foreground suppression, which can cause the suppression of 21-cm signal, and excess noise with small scales fluctuations in frequency, causing the loss of sensitivity and a measurement bias in the 21- cm signal power spectrum. \citet{2018MNRAS.478.1484G} have studied various wide field and calibration effects such as gain errors, polarized foregrounds, and ionospheric effects in power spectral analysis for LOFAR-LBA. They have also reported an excess power in the stokes I power spectrum, which might be due to incomplete sky-model or imperfect calibration.
In end-to-end simulations of full EoR power spectrum analysis, \citet{2016MNRAS.461.3135B} found that in the presence of an incomplete calibration catalog, the traditional per-frequency antenna calibration introduces contamination in the EoR window outside the wedge.
\citet{2017MNRAS.470.1849E} have also studied the impact of sky-based calibration errors for inaccurate sky models. Their work found that the unmodelled components of the foregrounds contaminate the EoR window by introducing a small frequency structure into gain solutions. The calibration errors associated with an incomplete sky model affect redundant calibration even in case of perfect redundancy and identical antenna beams, and these errors can exceed the predicted EoR signal \citep{2019ApJ...875...70B}. 
In case of redundancy calibration, errors in gain solutions are also introduced due to the non-redundancy of the arrays. \citet{2010MNRAS.408.1029L} shows that non-redundant baseline distributions result in spectral structures contaminating the EoR detections. \citet{2020MNRAS.499.5840D} have also studied the effect of non-redundancy on the gain solutions for a redundant array
such as HERA. This introduces characteristic patterns into the gain solutions, affecting the calibrated visibilities and power spectra. \citet{2018AJ....156..285J} investigated the effect of sky flux distribution and antenna position offsets in redundancy calibration. The position offsets introduce a bias into the complex gain solutions phase. They notice an enhancement in the bias as the distance between bright radio sources and the pointing center, and the flux density of the sources increases.
The deviations from perfect redundancy due to the antenna-to-antenna variations in redundant-baseline calibration produces considerable foreground power leakage from the wedge contaminating a considerable fraction of the EoR window \citep{2019MNRAS.487..537O}. The effect of primary beam non-redundancy using simulations is also studied in \citet{2021MNRAS.tmp.1553C}. They find that an additional temporal structure is induced in the gain solutions.

Usually, the calibration errors are relatively small, and the residual gain errors can be neglected for the bulk of the interferometric observations. However, for high dynamic range observations such as the redshifted 21-cm signal in the presence of strong foreground, accurate calibration of instrumental effects is of much need.

We investigate the effect of calibration inaccuracy by modeling the statistics of residual gain errors without referring to a particular mechanism that may lead to it. In a previous work \citep{2020MNRAS.495.3683K} (hereafter paper 1) we investigated the effect of time-correlated residual gain errors in the presence of large foreground through simulated observations. We find that residual gain error introduces a bias in the power spectrum measurements. This bias is significant for high dynamic range observations such as EoR observations. To demonstrate the effect of bias, we restricted ourselves to cases with no thermal noise. In practice, the presence of thermal noise limits the calibration accuracy. Furthermore, the residual gain errors can significantly contribute to the uncertainty in the power spectrum measurements. In this work, we include the thermal noise and use error propagation to find analytical expressions for the bias and variance of the power spectrum in the presence of strong foreground, residual gain errors, and noise. We compare and validate the analytical estimates with simulated observations and use the former to investigate various effects of the gain errors in power spectrum estimation. We provide a method to calculate the best possible estimate for the 21-cm power spectrum for a given interferometer and discuss the measures to be taken to enhance the chance of detection of the 21-cm signal.

The rest of the paper is organized in the following way. In section \ref{sec:section2}, we present an analytical framework to estimate the bias and variance of the power spectrum considering a simple model for the residual gain errors. In section \ref{sec:section3}, we compare the analytical estimates with simulated observations. Different effects of gain errors are investigated in section \ref{sec:section4}. We discuss the main results and conclude in section \ref{sec:section5}.

\section{Analytical estimates of Bias and Variance of Visibility correlation estimator}
\label{sec:section2}

This section finds an analytical estimate for the bias and variance in the power spectrum estimator for a given gain error model. Here, we restrict ourselves within a single frequency channel. The effect of frequency-dependent residual gain errors will be discussed in an accompanying paper.

In radio interferometric terms, the projected antenna separation vector in the plane of the sky, in units of observing wavelength, is called the `baseline vector'; we denote it with $\vec{U}$. We denote the spatial coherence function of the electric field of the radiation coming from the source as the `sky' visibility $\tilde{V }^S(\vec{U})$. We use the symbol ` $\tilde {} $' to denote the complex nature of the visibility and other quantities. In observations, a pair of antennas at a given time records the visibility function with modification by gains from different sources, including the gain of an individual antenna. The recorded signal includes measurement noise $\tilde{N}_i(\vec{U}_i)$. If a pair of antennas at a given time defines the baseline $\vec{U}_i(t)$, we may write the recorded visibility $ \tilde{V}(\vec{U}_i)$ at this baseline in terms of noise and gains as \citep{1996A&AS..117..137H, SynthesisImaging}
\begin{equation}
  \tilde{V}(\vec{U}_i) = \tilde{G}_i(t) \tilde{V}^S_i(\vec{U}_i) + \tilde{N}_i(\vec{U}_i),
  \label{eq:measurement}
\end{equation}
where $\tilde{G}_i(t) $ is the gain for the $i^{th}$ baseline. Note that, for a given pair of antennas, the projected antenna separation and hence the baseline vector changes with time as the antenna follows the source position in the sky. 

The gain $\tilde{G}_i(t)$ in each visibility arises from the individual gains of the antenna pairs used to estimate it. If the i$^{th}$  measurement of visibility at a time $t$ involves two antennae say $A$ and $B$, then the gain $\tilde{G}_i(t)$ can be written as 
\begin{equation}
 \tilde{G}_i(t)  = \langle \tilde{g}_A(t)  \tilde{g}^*_B(t)  \rangle, 
 \label{eq:gain}
\end{equation}
where $\tilde{g}_A$ and $\tilde{g}_B$ are the gains of the individual antenna and the angle brackets represent the average over the integration time.

Interferometric noise $\tilde{N}_i(\vec{U}_i)$ can be considered Gaussian random with zero mean. Given an antenna characteristics and its source equivalent flux density (SEFD), frequency width of the channel $\Delta \nu$ and integration time for each visibility $\Delta \tau$, the standard deviation of the noise in the real or imaginary part of each visibility can be written as \citep{1986isra.book.....T}
\begin{equation}
    \sigma_N(\vec{U}) = \frac{\text{SEFD}}{\sqrt{2\Delta \nu\,  \Delta \tau}}.
    \label{eq:visnoise}
\end{equation}
It is safe to assume that the interferometric noise is uncorrelated in time. Hence, its auto-correlation functions are zero except at zero delay.

\subsection{Power Spectrum Estimator}
A widely used statistical property of the sky brightness distribution is its power spectrum (\citet{2001JApA...22..293B, 1995A&A...293..507L} etc.). As the redshifted 21-cm signal is expected to be faint and hard to detect with imaging, estimating its power spectrum or equivalently intensity mapping gives a possible probe of the evolution of the baryonic matter distribution over cosmic time. \citet{2001JApA...22..293B} shows that visibility correlation directly measures the power spectrum. This method and its variants (\citet{2007MNRAS.378..119D, 2014MNRAS.445.4351C, 2016MNRAS.463.4093C, 2019MNRAS.483.3910C, 2019MNRAS.483.5694B} etc.) have been used to estimate the angular power spectrum of the diffused galactic foreground \citep{2012MNRAS.426.3295G, 2017MNRAS.470L..11C, 2019MNRAS.487.4102C, 2020MNRAS.494.1936C} as well as the power spectrum of \HI distribution in nearby galaxies \citep{2009MNRAS.398..887D, 2013MNRAS.436L..49D, 2020MNRAS.496.1803N}. These works propagate the uncertainties in each visibility estimate and combine that with the sample variance error in measuring the power spectrum to quote uncertainties in the power spectrum estimates. 
In this work, we use the estimator discussed in \citet{2014MNRAS.445.4351C}, where visibilities are gridded before estimating the power spectrum. Given an angular field of view of $\theta_0$ to which the telescope is sensitive, it has been shown \citep{2001JApA...22..293B, 2005MNRAS.356.1519B, 2014MNRAS.445.4351C} that the visibilities in the nearby baselines remains correlated to a baseline separation of $\Delta U < \frac{1}{\pi \theta_0}$. The size of the uv-grids is chosen such that they are large enough to include a sufficient number of baselines in a given uv-grid and small enough to have all visibilities in the uv-grid correlated. In each uv-grid, they estimate the power spectrum by correlating visibilities only in nearby baselines, omitting the visibility auto-correlations. This drastically reduces the noise bias in estimates of the power spectrum in uv-grids. The contribution from each uv-grid within a given annulus in $U = \mid \vec{U} \mid $ is then combined, and the real part of it is used to quote the value of the isotropic power spectrum for the baseline separation $U$. We may schematically write it as
\begin{equation}
    \mathcal{E} \left \{ P( U ) \right \}  = \mathcal {R} \left [\langle   \tilde{V}(\vec{U})^*  \tilde{V}(\vec{U}+\Delta \vec{U})  \rangle \right ].
    \label{eq:defps}
\end{equation}
Here, the average is taken over the uv-grid first and then within the annulus, as explained above. Note that the power spectrum estimator here assumes that a perfect calibration is done and the gains are all unity. In such a case, the power spectrum estimate has no bias arising from instrumental noise, and its uncertainties can be written as  \citep{2008MNRAS.385.2166A, 2011arXiv1102.4419D}
\begin{equation}
  \sigma_P^2 = \frac{P^2(U)}{N_G} + 2\frac{P(U)\sigma_N^2}{N_B} + 2\frac{\sigma_N^4}{N_B},
  \label{eq:psvarng}
\end{equation}
where $N_G$ is the number of independent estimates of the power spectrum in a given annulus bin at $U$, $N_B$ is the total number of visibility pairs in the bin. 

Using a numerical simulation of observed visibilities with a model of residual gain errors in paper 1, we have shown that the latter introduces bias in the power spectrum estimates. Here we use an improved model for the gain error and provide an analytical estimate for both bias and variance of the power spectrum in the presence of residual gain errors and noise for the power spectrum estimator discussed above. These analytical expressions are then compared with simulated observations.

\subsection{Effect of gain errors in visibility correlation}
\label{subsec:2.2}
As discussed above, correlating the observed visibilities at nearby baselines gives an unbiased estimate of the power spectrum in the presence of no residual gain errors. However, in the presence of gain errors, the estimator discussed above is biased, and the uncertainties in the estimates of the power spectrum increase. This is particularly important while observing a field that requires a high dynamic range in sensitivity. To understand the additional effect from gain errors, we discuss how we can get a pair of nearby baselines in a given uv-grid where the visibilities are measured and correlated. Note that every visibility is measured by correlating the electric fields from a pair of antennas. The visibility is measured at a baseline $\vec{U}$ given by the antenna separation projected in the plane of the sky, weighted by the inverse of the observing wavelength. As the sky rotates with respect to the observer, the same pair of antenna gives rise to different measurements of the visibilities at different baselines $\vec{U}$. Hence, for a given antenna pairs A and B, at a time t, the visibility is measured at a baseline $\vec{U}_{AB}(t)$, we write this visibility as $\tilde{V}_{AB}(t)$. For nearby baseline correlation, we can get a pair of baselines in the following ways:
\begin{itemize}
    \item {\bf Type 1} Correlation of the visibilities measured by same antenna pair, at different time, i.e.
      $\langle  \tilde{V}_{AB}(t) \tilde{V}^*_{AB}(t')\rangle$
    \item {\bf Type 2} Correlation of the visibilities measured by antenna pairs having one antenna in common, measured at the same time, i.e.
    $\langle  \tilde{V}_{AB}(t) \tilde{V}^*_{AC}(t)\rangle$
    \item {\bf Type 3} Correlation of the visibilities measured by antenna pairs having one antenna in common measured at different times.
     $\langle  \tilde{V}_{AB}(t) \tilde{V}^*_{AC}(\ t')\rangle$  
    \item {\bf Type 4} Correlation of the visibilities measured by antenna pairs having no antenna in common measured at any time.
     $\langle  \tilde{V}_{AB}(t) \tilde{V}^*_{CD}(t')\rangle$.
\end{itemize}
Note that the noise is uncorrelated between any two measurements considered for all the above cases. If we consider any uv-grid in the baseline plane in which all nearby baseline correlations are performed to estimate the power spectrum, then those baseline pairs may have a contribution from all four cases discussed here. As gain in the interferometer depends on the antenna, different types of baseline pairs contribute differently to the excess gain and uncertainties of the visibility correlations. We define the fraction of baseline pairs of type `i' (where i can be any of 1-4 above) as $n_i (\vec{U_g})$ in a uv-grid. Clearly, this depends on the baseline of the uv-grid $\vec{U_g}$ through the antenna configuration of the telescope and source position for observation.

\subsection{Modelling the residual Gain errors}
\label{subsec:Gmodel}
The first step to preparing the observed data for scientific purposes is to estimate the antenna gains using primary calibration or self-calibration methods. In the simulation, under perfect calibration, all gains are unity. In practice, interferometric calibration is affected by the non-zero noise in the system, the time-dependent ionospheric variation, the calibration algorithm used, etc. Here we assume that the best calibration procedure for the data in consideration has been applied, and only residual gains for each antenna contribute to the gain term $\tilde{G}_i(t)$. In such a case, we can write the gain from an individual antenna (say antenna A) as
\begin{equation}
 \tilde{g}_A(t) = \left [ 1+\delta_{AR}(t) + i\delta_{AI}(t)\right ]
 \label{eq:gaindef}
\end{equation}
where $\delta_{AR}(t)$ and $\delta_{AI}(t) $ stands for the real and imaginary part of the residual gain error. As the best possible calibration is already performed, we assume that the residual gains are small compared to unity and zero mean random numbers. Here we do not consider any frequency dependence of the gain and its polarisation properties. This work assumes that the residual gain errors are Gaussian random and quantifies them with their variance and two-point correlation functions. \footnote{Note that the residual gain errors may have non-zero higher-order correlations and hence can have non-Gaussian properties. However, we expect the two points correlations to dominate here.} We further assume that the real and imaginary parts of the gain from a given antenna and residual gains from the different antenna are not correlated. These assumptions work fairly well in a real observation; however, there can be departure coming from unidentified low-level radio frequency interference, correlated structural change of the antenna with elevation, etc. All these effects change with observatories. These effects can be included in the framework we discuss here if needed. We denote the variance $\sigma^2_{AC}$ and the normalized two-point correlation $ \xi_{AC}(\tau)$ of the residual gain from antenna A as
\begin{equation}
\sigma^2_{AC} = \langle \delta^2_{AC} \rangle, \ \ \ \ \xi_{AC}(\tau) = \langle \delta_{AC}(t) \delta_{AC}(t+\tau) \rangle / \sigma_{AC}^2,
 \label{eq:gainmod0}
\end{equation}
where $C$ is to be read as $R$ or $I$ for real and imaginary parts of the gain and hence $\delta_{AC}$ can be $\delta_{AR}$ or $\delta_{AI}$ for the real or imaginary part of the gain respectively. The two-point correlation of the residual gains is a function of (time) delay $\tau$ and is normalized such that its value is unity at zero delay. Given an observation, it is possible to estimate these properties of the gain from the calibration solutions. Here, for simplicity, we assume that the variance for all the antennas in the array is the same. We shall denote these as $\sigma_R^2$ and $\sigma_I^2$. We further assume that the normalized two-point correlation functions for all antennae are the same irrespective of them being of real or imaginary parts; we denote this by $\xi(\tau)$.

The two-point correlation of the gain is expected to have contributions from various sources, including sky model errors, temperature fluctuations, instrumental beam variation, electronic gain variations, etc. Here, we assume that the best calibration possible is already performed. We are interested in the residual gains only, and the two-point correlation function $\xi(\tau)$ represents the two-point correlation in the residual gain.
As the gain is calibrated at a given time interval, the calibration procedure is expected to reduce time correlation in the residual gain. However, gain errors at smaller time scales as well as long-term effects over multiple days can still be present in the residual gains. For the purpose of this work, we assume that the normalized two-point correlation function $\xi(\tau)$ of the residual gain error is unity at zero delay $\tau$ and model it as
\begin{equation}
\xi(\tau) = \exp \left [ - \frac{\tau^2}{2 T_{corr}^2}\right ]
\label{eq:xidef}
\end{equation}
where $T_{corr}$ gives the correlation time of the residual gains. Note that the above model considers that the residual gain errors have only short time scale variations. In reality, the two-point correlation function of the residual gain error can be rather complicated. In practice, it needs to be estimated for any observation and used to calculate the effect of residual gain for the given observation.

\subsection{Bias and Variance of the power spectrum}
\label{subsec:AnalyticBP}
The formalism to include the effect of residual gain errors in the bias and variance of the power spectrum is rather general and can be used for any high dynamic range observations. Here we are particularly interested in the case of observing the power spectrum of redshifted 21-cm brightness fluctuations from the EoR. An additional complication in observing the redshifted 21-cm signal is the presence of emission from other sources in the observing frequency. This includes radiation from the compact extragalactic sources as well as the diffuse galactic synchrotron radiation. These signals are collectively known as foregrounds. We may write the sky visibility $\tilde{V}^{S}(\vec{U}_i)$ as a combination of the redshifted 21-cm signal $\tilde{V}^{HI}(\vec{U}_i)$ and the foreground $\tilde{V}^{F}(\vec{U}_i)$ as
\begin{equation}
\tilde{V}^{S}(\vec{U}_i) = \tilde{V}^{HI}(\vec{U}_i) + \tilde{V}^{F}(\vec{U}_i).
\label{eq:defHF}
\end{equation}
The properties of the foregrounds are well studied and measured \citep{2002ApJ...564..576D, 2008MNRAS.385.2166A, 2008MNRAS.389.1319J, 2012MNRAS.426.3295G, 2019MNRAS.487.4102C} etc. The power spectrum of the foreground is known to be six to seven orders of magnitude higher amplitude than that of the redshifted 21-cm signal \citep{1999A&A...345..380S, 2002ApJ...564..576D}. Hence, observation of the 21-cm power spectrum requires high dynamic range calibration. There are several methods adopted in literature (\citep{2010MNRAS.409.1647J, 2017NewA...57...94C, 2021MNRAS.500.2264H} etc.) to estimate the foreground and mitigate its effect. Here we assume that the foreground is well estimated and subtracted from the observed and calibrated visibilities. These foreground subtracted visibilities are then used for estimating the redshifted 21-cm power spectrum. The excess bias and variance in the power spectrum in the presence of the residual gain errors is a combined effect of the gain errors and foregrounds \citep{2020MNRAS.495.3683K}.

As discussed earlier, the power spectrum estimator discussed here is unbiased in the absence of residual gain errors. We define the bias in the power spectrum estimator as the difference of the mean value of the right-hand side of equation \ref{eq:defps} with and without residual gain errors. \citet{2008MNRAS.385.2166A, 2011arXiv1102.4419D} discuss the calculation of uncertainty in the power spectrum estimates by the propagation of noise in visibilities. Here, we adopt a similar method to calculate the uncertainty in the power spectrum in the presence of residual gain errors as the variance of the visibility correlation with the propagation of errors coming through the noise and residual gains.

In the presence of residual gain errors, the power spectrum estimated from visibility correlation would be biased, and its uncertainty will have an additional contribution from the residual gain errors. However, in the context of 21-cm emission, if $\tilde{V}^{S}(\vec{U}_i)$ has contribution only from the 21-cm emission since the residual gain errors are expected to be much smaller than unity, the bias in the power spectrum is rather small and can be ignored. The effect of the residual gain errors manifests itself in the presence of a strong foreground signal. Here, we assume that a good estimation of the foreground signal already exists, and a foreground subtraction has been performed to extract the 21-cm signal from the observed visibilities. If these visibilities are now used to estimate the 21-cm power spectrum, the estimates will have bias and increase uncertainties resulting from the residual gain errors and foreground signal.

We denote the bias in the power spectrum estimate as $\mathcal{B}_P (U)$ and its variance as $ \sigma^2_{P}(U) $. In what follows, we outline the analytical steps to calculate the bias $\mathcal{B}_P(U)$ and variance $\sigma_P^2(U)$ of the 21-cm power spectrum measurements in the presence of strong foreground and residual gain errors. The various assumptions done in this calculation are kept in the italic font for quick access by the reader. Note that we use a foreground subtraction approach here \citep{2006ApJ...648..767M, 2009ApJ...695..183B, 2017NewA...57...94C}, where the power spectrum is calculated by correlating the visibilities in the same observed frequency channel. Though the foreground avoidance method is expected to reduce the effect of foreground drastically from the visibility correlation \citep{2010ApJ...724..526D, 2012ApJ...752..137M, 2012ApJ...745..176V}, it is important to observe that \HI signal correlation at different frequency channels also reduce the \HI signal \citep{2003JApA...24...23B, 2005MNRAS.356.1519B}.

Furthermore, a residual bandpass component, like the residual time-dependent gain considered here, will leak the foreground to the outer part of the wedge as well. Due to the bandpass calibration uncertainties, the correlated sky signal residuals will remain in the data, which will lead to power in the 2-D power spectrum contaminating the EoR window \citep{2016PASA...33...19T, 2016MNRAS.461.3135B, 2012ApJ...752..137M}. These effects of residual frequency-dependent gains are being investigated and will be presented in the next iteration of the paper in this series.

The recorded visibility is expressed in terms of the `sky' visibility, gain $ \tilde{G}_i(t)$ and correlator noise in eqn \ref{eq:measurement}. Expressing the gain in terms of the antenna based gains using eqn \ref{eq:gain}, we write
\begin{equation}
\tilde{G}_i = \langle \tilde{g}_A(t) \tilde{g}^*_B(t) \rangle = 1 + \tilde{G}_i^R,
\end{equation}
where $\tilde{G}_i^R$ can be interpreted as the excess over unity in the residual gain.
Here the angle brackets represent the average over the integration time for each visibility measurements. We consider the following assumptions\\
{\it Assumption I: Antenna gains from the different antennas are uncorrelated,\\
Assumption II: Real and imaginary parts of the gains are uncorrelated.}

For the compact source foreground, it is in principle possible to build up a good sky model over largely repeated observations at a given direction in the sky. However, for the diffuse galactic synchrotron emission, it may be challenging to build up such a sky model with interferometric observations, and an alternative approach may be necessary \citep{2008MNRAS.389.1319J}. In this work, we deal with a power spectrum estimation that assumes the preexistence of a sky visibility model and uses foreground subtraction. Hence, \\
{\it Assumption III: We have a known model for the foreground sky visibilities.}
 
Hence, the residual visibilities $\tilde{V}^R_i$ from the $i^{th}$ baseline, after subtraction of the known foreground components, are given by
\begin{equation}
    \tilde{V}^R_i = \tilde{V}^{HI}_i + \tilde{G}_i^R \tilde{V}^S_i + \tilde{N}_i,
\end{equation}
where $\tilde{V}^{HI}_i$ is the redshifted 21-cm signal. Correlating the residual visibilities, in principle, gives the \HI power spectrum $P$. Owing to the finite beam of the antenna, the nearby `sky' visibilities remain correlated within a baseline region of $1/(\pi \theta_0)$. The power spectrum estimator we use here \citep{2014MNRAS.445.4351C} first grid the visibilities in baseline grids and estimate visibility correlation in each grid. To avoid noise bias, here we exclude the visibility auto-correlations \citep{2005MNRAS.356.1519B}. Estimate of the power spectrum $P'_g$ over a grid can be written as
\begin{equation}
    P'_g = \mathcal{R} [ \langle \tilde{V}^R_i \tilde{V}_j^{R'*} \rangle_g ],
 \end{equation}
where $\mathcal{R} [\  ]$ denotes the real part of the visibility correlation, $i, j$ denotes two different baselines and $\langle \ \rangle_g$ denotes average over the grid. Here we further assume the followings:\\
{\it Assumption IV: Noise in different baselines are uncorrelated.\\
Assumption V: The sky signal, antenna gains and noise have no cross correlations.}\\

Hence, the power spectrum estimates in each grid can be written as
\begin{equation} 
 P'_g = \langle \tilde{V}_i^{HI} \tilde{V}_j^{HI*} \rangle_g +  \langle \tilde{G}_i^R \tilde{G}_j^{R'*} \rangle_g \langle \tilde{V}_i^S \tilde{V}_j^{S*} \rangle_g.
\end{equation}
Note that the sky visibilities $\tilde{V}_i^S$ contain both the redshifted 21-cm signal as well as the foreground. We can use the following assumptions to simplify this expression further.\\
{\it Assumption VI: Foreground and redshifted 21-cm signals are uncorrelated.\\
Assumption VII: We can neglect the contribution from the 21-cm signal from $\tilde{V}_i^S$ in the last term above.}\\

\subsection*{Bias}
 The bias $\mathcal{B}_{Pg}$ in the estimate of the power spectrum in a grid can be written as
\begin{equation} 
\mathcal{B}_{Pg} = P'_g - P_g = \langle \tilde{G}_i^R \tilde{G}_j^{R'*} \rangle_g P_{Fg},
\end{equation}
where $P_g$ and $P_{Fg}$ are the redshifted 21-cm and foreground power spectrum in the grid. As mentioned in the section \ref{subsec:2.2}, in each grid, four types of baseline pairs contribute to the visibility correlations. To calculate the contribution from the different baseline pairs in a grid, we need to estimate $\langle \tilde{G}_i^R \tilde{G}_j^{R'*} \rangle_g$. Here we show the calculation for the baseline pair of Type I. In this type the pair of baselines contributing in the visibility correlation originates from the same antenna pairs but at different time. The quantity $\tilde{G}_i^R \tilde{G}_j^{R'*}$ for a particular baseline pairs $i,j$ constructed by antenna pairs $A,B$ can be written as
\begin{eqnarray}
& \tilde{G}_i^R (t) \tilde{G}_j^{R'*}(t') = \langle \delta_{AR} (t) \delta_{AR} (t') \rangle \\ \nonumber
& + \langle  \delta_{BR} (t) \delta_{BR} (t') \rangle +  \langle \delta_{AI} (t) \delta_{AI} (t') \rangle + \langle \delta_{BI} (t) \delta_{BI} (t') \rangle,
\end{eqnarray}
where we have used assumptions I and II. If we assume that there are a total of $N_g$ number of visibility correlations in a grid with $N_{g1}$ giving the visibility pairs of Type 1, the contribution to $\langle \tilde{G}_i^R (t) \tilde{G}_j^{R'*}(t') \rangle_g$ from the baseline pairs of type 1 can be written as
\begin{equation}
\langle \tilde{G}_i^R (t) \tilde{G}_j^{R'*}(t') \rangle_{g1} = 2\ n_1\ \frac{ \sigma_R^2 + \sigma_I^2  }{N_{g1}} \sum_{A=1}^{N_{g1}} \xi_{AR},
\label{eq:eqn16}
\end{equation}
where $n_1 = N_{g1}/N_g$ is the baseline pair fraction of type 1 and we have used the definitions of $\sigma_{AR}^2$ and $\xi_{AR}$ from eqn \ref{eq:gainmod0}. 
The above eqn~\ref{eq:eqn16} is obtained using the following additional assumptions:\\
{\it Assumption VIII: Statistical properties of the gains i.e. $\sigma_R$ and $\xi_R$, for all antenna are similar.}\\
Assumption IX: Gain correlation function $\xi$ is the same for real and imaginary parts of the gain, though the variances can be different.\\
Since each antenna here contributes a factor of $\xi$ and we assume all antennae have similar statistical properties, a factor of $2$ for every baseline pair is coming in eqn~\ref{eq:eqn16}.

Note that the angular brackets $\langle \ \rangle$ used above denotes average over integration time over one visibility measurement, whereas $\langle \ \rangle_g$ is considered as the average over baseline pairs of type 1 in a particular grid. A pair of antenna traces an ellipse in the baseline plane as time progresses. Part of this ellipse would trace through a particular grid giving rise to different baselines arising from the same antenna pairs. This gives rise to the next simplifying assumption in our calculation:\\
{\it Assumption X: If the time for the baseline to trace the arc in a grid $T_D$ is much larger than the integration time $\Delta \tau$, then we can approximate }
\begin{equation}
 \chi(U) = \frac{1}{N_{g1}} \sum_{A=1}^{N_{g1}} \xi_{AR}  
 \end{equation}
The quantity $\chi(U)$ gives the integrated effect of the time-correlated residual gain errors over the uv-grids. If the correlation time of the residual gain errors $T_{corr}$ are larger than the integration time $\Delta \tau$ then
\begin{equation}
    \chi(U) = \frac{1}{T_D^2}\int_{\Delta \tau}^{T_D} (T_D-\tau)\xi(\tau)d\tau.
    \label{eq:chi}
\end{equation}
Here $T_D = \frac{\Delta U \times T_{24}}{2\pi U}$ is the time taken by baseline track of an antenna pair to cross a uv-grid of size $\Delta U$ at baseline $U$. $T_{24}$ corresponds to one sidereal day.
Hence, the contribution to $\langle \tilde{G}_i^R (t) \tilde{G}_j^{R'*}(t') \rangle_g$ from the baseline pairs of type 1 is
\begin{equation}
\langle \tilde{G}_i^R (t) \tilde{G}_j^{R'*}(t') \rangle_{g1} = 2\ n_1\ (\sigma_R^2 + \sigma_I^2)   \ \chi.
\end{equation}
Considering contribution from the other three baseline pairs in a similar way we can write for a given day of observation for a given baseline grid
 \begin{equation}
     \langle \tilde{G}_i^R \tilde{G}_j^{R'*} \rangle_g = \left [(2n_1+n_3)\chi + n_2 \right ]\ (\sigma_R^2 + \sigma_I^2).
      \end{equation}
Considering the grids to have independent estimates of the power spectrum and a further assumption that\\
{\it Assumption XI: The gain errors do not have any long term correlation}\\
we use an azimuthal average around a certain baseline $U$ to estimate the power spectrum. This then, gives a bias in the power spectrum estimate as
 \begin{equation}
     \mathcal{B}_P = \left [(2n_1+n_3)\chi + n_2 \right ]\frac{(\sigma_R^2 + \sigma_I^2) }{N_d} P_F,
     \label{eq:BiasA}
 \end{equation}
 where $N_d$ is the number of days of observation and $P_F$ is the foreground power spectrum.
 
\subsection*{Variance}
We first calculate the variance of the estimator in a grid, where
\begin{equation}
     {\sigma^2_{Pg}} = P'_{g2} - \langle P'_g \rangle_g ^2,
 \end{equation}
 where $P_{g2}$ is given as
 \begin{equation}
     P'_{g2} = \frac{1}{4} \langle [ \tilde{V}_i^R \tilde{V}_j^{R'*}  +  \tilde{V}_i^{R*} \tilde{V}_j^{R'}  ]^2 \rangle_g.
 \end{equation}
Note that the quantities $\tilde{V}_i^R$ depend on the gain errors. In calculating the four-point functions above, we do a further simplifying assumption:\\
{\it Assumption XI: The gain errors are Gaussian random variables.}\\
 We follow a similar procedure to calculate the variance in a grid as like for the variance for $N_d$ days of observations. Since the grid size is chosen in a way that in an annulus in baseline $U$ the estimates of the power spectrum from different grids remains uncorrelated, the variance in the power spectrum estimate can be written as
 \begin{equation}
  \sigma^2_{P} = \frac{1}{N_G} \sum_{g=1}^{N_G}\sigma_{Pg}^2,
 \end{equation}
 where $N_G$ is the total number of grid points in an annulus. Writing the total number of baseline pairs to estimate the power spectrum in a given annulus as $N_B$, the analytical expression for the variance in \HI power spectrum estimates in an annulus is
\begin{eqnarray}
\nonumber
  \sigma^2_{P} &=& \left[\frac{P^2_{HI}}{N_G} + \frac{2\sigma_N^2 P_{HI}}{N_B N_d} + \frac{2\sigma_N^4}{N_B N_d^2} \right]+ 
   \left [ \frac{4 \sigma_N^2 (\sigma_R^2 + \sigma_I^2)  P_F}{N_B N_d^2}\right] 
   \\ \nonumber
   &+& \bigg [ \left [(4n_1^2 + n_3^2) \chi^2 + n_2^2 \right ]\left [ 3\sigma_R^4 + 3\sigma_I^4 + 2 \sigma_R^2 \sigma_I^2 \right ] 
   \\ 
   &+& 8 (\sigma_R^4 + \sigma_I^4)
     \frac{4\sigma_N^2 P_F^2}{N_G N_d^2} \bigg ]
    \label{eq:VarA}
\end{eqnarray}
Here $P_F$ is the power spectrum of the foreground signal. We assume that the power spectrum of the foreground is much higher than that of the redshifted 21-cm signal. In reality, there is a contribution to bias from the 21-cm signal as well; however, they are much smaller than the signal itself and hence are ignored here. The quantities $N_G$ and $N_B$ are the number of independent estimates of the power spectrum and the total number of visibility correlations in a baseline annulus. We assume here that the noise and residual gain errors are not correlated between the observations done on different days; $ N_d$ is the number of days of observation. Note that both the bias and variance change with baseline, and apart from $\sigma_N, \sigma_R$ and $\sigma_I$ all the other factors in the expressions for bias and variance are baseline dependent. The term $\frac{P^2_{HI}}{N_G}$ is the contribution from the sample variance and decreases with the baseline. In all our investigations presented here, this term is negligible.

For convenience, we define

\begin{eqnarray}
\text{Term 1} &=& \left [ \frac{P^2_{HI}}{N_G} + \frac{2  \sigma_N^2 P_{HI} }{N_B N_d} + \frac{2 \sigma_N^4}{N_B N_d^2} \right ]
\\ \nonumber
\text{Term 2} &=& \left [ \frac{4 \sigma_N^2 (\sigma_R^2 + \sigma_I^2)  P_F}{N_B N_d^2}  \right] 
  \\ \nonumber
\text{Term 3} &=& \bigg [ \left [(4n_1^2 + n_3^2) \chi^2 + n_2^2 \right ]\left [ 3\sigma_R^4 + 3\sigma_I^4 + 2 \sigma_R^2 \sigma_I^2 \right ] 
   \\ 
   &+& 8 (\sigma_R^4 + \sigma_I^4)
     \frac{4\sigma_N^2 P_F^2}{N_G N_d^2} \bigg ]
\end{eqnarray}

As expected, the power spectrum estimator is unbiased in the absence of gain errors. Hence, the `Term 1' in variance is independent of the gain errors. The `Term 2' and `Term 3' represent the contribution in the variance of the power spectrum arising because of the presence of residual gain errors and depend on residual gain errors through both ($\sigma_R, \sigma_I$) and $\chi$. The function $\chi$ is non zero if the correlation time of the residual gain errors $T_{corr}$ are larger than the integration time $\Delta \tau$. If the gain errors do not have any time correlation, $\chi(U)$ is zero, and the contribution to bias and variance originates from $\sigma_R$ and $\sigma_I$ only. In the presence of time-correlated residual gain errors, the function $\chi(U)$ needs to be calculated additionally.
The contribution from the time-correlated gain errors also comes through the baseline pairs of type $1$ and $3$. The baseline pair of type $4$ does not contribute to the bias and variance. This is expected as this type has four different antennas, and we have assumed that the residual gain errors in the different antennas are uncorrelated. We observe that if the gain errors are not correlated in time, that is $T_{corr}=0$, the residual gain errors still introduce bias and excess variance in the power spectrum estimates. In the calculation shown here, we have neglected terms higher than the fourth power in the standard deviation of residual gain errors.

\section{Comparing analytical expression of bias and variance with simulation}
\label{sec:section3}
Here we study the effect of residual gain errors in estimating the redshifted 21-cm power spectrum in the presence of strong foreground. As has been shown analytically, residual gain errors introduce a bias and enhance the variance of the power spectrum. The bias and the variance of the power spectrum depend on the gain error model, baseline configuration, and the foreground model. This section compares the analytical results against simulated observations with a given gain error and foreground model.

\begin{figure}
    \centering
    \includegraphics[width=0.5\textwidth]{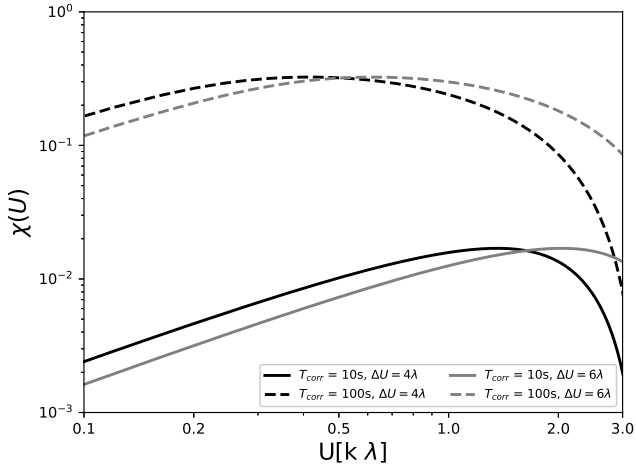}
    \caption{Variation of $\chi$ with baseline U for different correlation times $T_{corr}$ and uv-grid size $\Delta U$ for an integration time of $16$ seconds. }
    \label{fig:Chi}
\end{figure}

\begin{figure}
    \centering
    \includegraphics[width=0.5\textwidth]{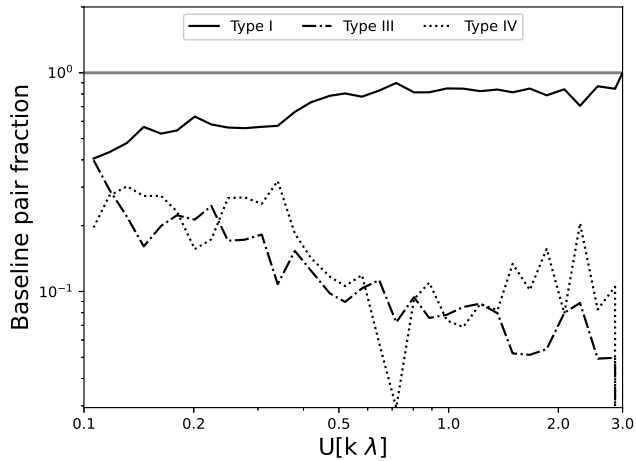}
    \caption{Variation of baseline pair fractions with baseline U for GMRT, 8 hour of observation. Integration time is 16 sec and uv-grid size is $0.004$ $k\lambda$. Fraction of baseline pairs of type II is zero for above used parameters for GMRT.}
    \label{fig:n_i}
\end{figure}

We use the Upgraded Giant Metrewave Radio Telescope (uGMRT) \footnote{The upgraded GMRT: opening new windows on the radio Universe  \citep{2017CSci..113..707G}} baseline configuration for simulating the effect of gain errors. The uGMRT has thirty fully steerable $45$ meter diameter parabolic dishes spread over a region with a maximum antenna separation of $25$ kilometer. We set up the observation at a central frequency of $150$ MHz, where the telescope gives a circular field of view with a radius of $\theta_0 = 186^{'}$ and a maximum baseline of $12.5$ k$\lambda$. As we are interested in the effect of gain in a single frequency channel here, we choose a channel width of $62.5$ kHz. Given that the system temperature at this frequency is dominated by its contribution from the sky, the above configuration gives $\sigma_N = 1.36$ Jy of noise per visibility for $15$ sec integration time. 

The major modification in the bias and variance of the power spectrum arises from the strong foreground. Here we use a point source foreground model based on the differential source count at $150$ MHz estimated in \citet{2017A&A...598A..78I}. We use this differential source count to generate a point source sky model. Paper 1 gives a detailed description of the methodology followed to generate the point source sky model, and we follow a similar procedure here. To consider the effect of sample variance, we generate $128$ realizations of the point source sky models. All the models have $7250$ sources with flux density ranging between $300$ mJy to $1$ Jy over a radius of $200^{'}$. The redshifted 21-cm power spectrum at $150$ MHz is $10^6-10^7$ times smaller than the foreground and provides a minimal effect on the bias and variance of its power spectrum in the presence of residual gain errors. We do not include the 21-cm signal in our simulation. To consider a specific baseline distribution, we choose the declination of the center of our simulated point source field at $+30^{\circ}$ and a total observation time of eight hours symmetrically distributed from the transit time of the source. This configuration is also used to estimate the baseline pair fractions for the analytical calculation of the bias and variance of the power spectrum. Figure~\ref{fig:Chi} show the variation of $\chi$ as a function of baseline for different combinations of $T_{corr}$ and uv-grid size $\Delta U$. 
For all choices of $\Delta U$ and $T_{corr}$, the effect of $\chi$ is less at shorter baselines and is effectively negligible at longer baselines. Larger correlation time increases the amplitude of $\chi$ and hence its contribution to the bias and variance of the power spectrum. Since the limits in the integration for $\chi$ depend on the uv-grid size, we also see that a larger uv-grid size keeps the value of $\chi$ at a significant level for a longer baseline value. This suggests a gain calibration with a lower correlation time for residual gain, and smaller uv-grid size is preferred to reduce the bias and variance of the power spectrum.

We choose a uv-grid size $\Delta U = 4\ \lambda$ for further calculations. This is smaller than $1/(\pi \theta_0) = 6\ \lambda$ and large enough to accommodate enough baseline pairs in the majority of the uv-grids. 
Figure~\ref{fig:n_i} show the variation of the baseline pair fractions as a function of $U$. We observe that the uGMRT configuration provides mainly the baseline pairs of type I for visibility correlation, whereas the baseline pair of type II is completely absent. As we will see, the baseline pair fractions have a significant role in deciding the strength of the bias and excess variance in the power spectrum. We also note here that the baseline pair fractions are expected to be significantly different for different telescopes.

\citet{2005MNRAS.356.1519B} has estimated the expected redshifted 21-cm power spectrum at different observing frequencies. They find that at $150$ MHz, the power spectrum remains almost constant to about $0.5$ k$\lambda$, and its amplitude reduces drastically beyond $2$ k$\lambda$. Hence, we show the baseline pair fractions up to $3$ k$\lambda$ only. We observe here that $n_2 \sim 0$ and baseline pair fraction of type $1$ dominates for all baselines. 

\begin{figure}
    \centering
    \includegraphics[width=0.5\textwidth]{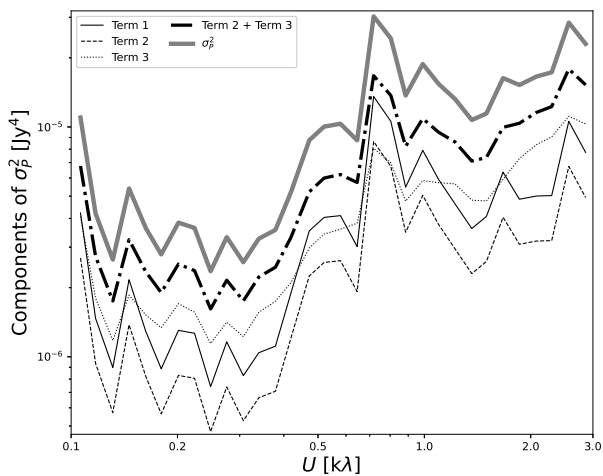}
    \caption{Comparison of contribution from different terms in the expression for variance of the power spectrum $\sigma_P^2$. The result shown here is for $8$ hours of the uGMRT observation with an integration time of $16$ sec and bandwidth of $256$ kHz. We have chosen $5\%$ gain error in both real and imaginary parts of the visibilities and $T_{corr} = 16$ seconds. }
    \label{fig:sigcomp}
\end{figure}

\begin{figure*}
    \centering
    \includegraphics[width=1.0\textwidth]{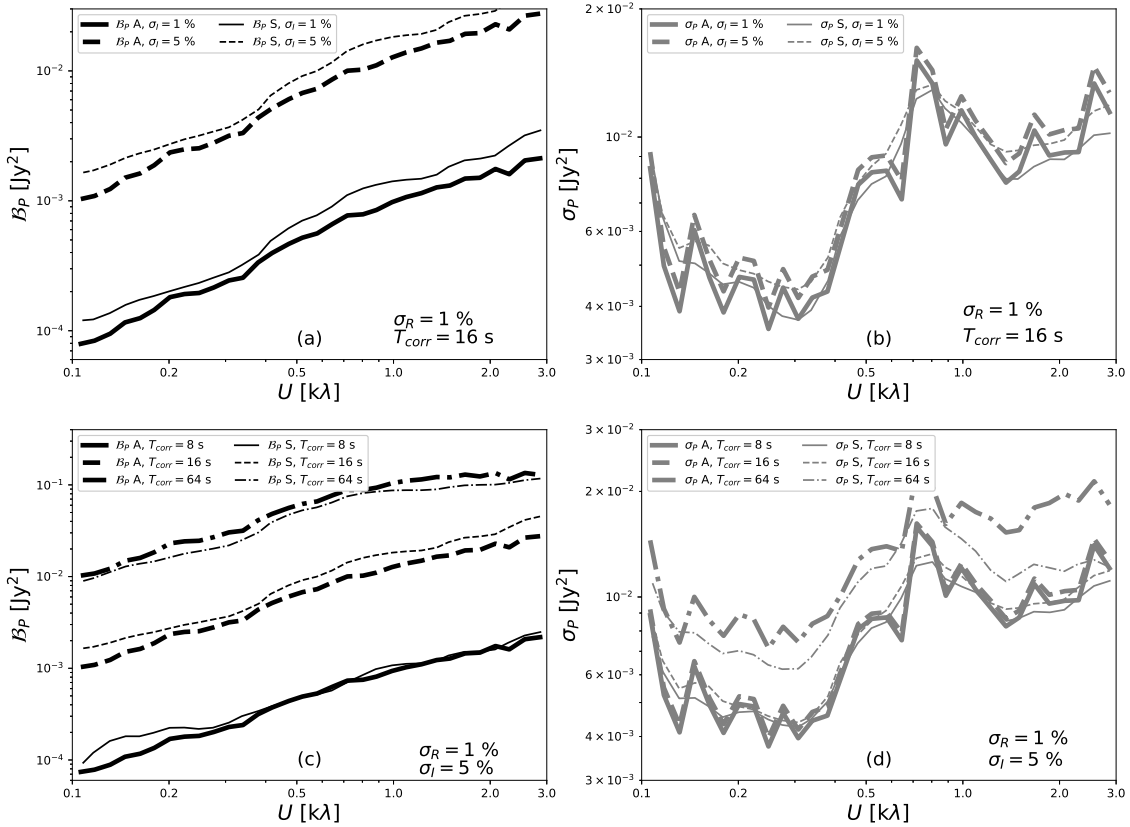}
    \caption{ We compare the analytical estimates of $\mathcal{B}_P$ (thick black) and $\sigma_P$ (thick grey) with their estimate from simulation (thin black and thin grey respectively for $\mathcal{B}_P$ and $\sigma_P$). In the left side we show the comparison plots for $\mathcal{B}_P$ and the comparison plots for $\sigma_P$ is shown in the right side. The plots are shown for eight hours of observing time and a bandwidth of $64$ kHz. In the top panel we keep $\sigma_R = 1$ \% and $T_{corr} = 16$ sec and show $\mathcal{B}_P$ in (a) and $\sigma_P$ in (b), for different values of $\sigma_I$. In the bottom panel we fix $\sigma_R = 1$ \% and $ \sigma_I = 5$ \% and show the bias $\mathcal{B}_P$ in (c) and $\sigma_P$ in (d), for $T_{corr} = 8, 16, 64$ seconds.}
    
    \label{fig:gcomp1}
\end{figure*}

The gain error model we discussed earlier can be expressed in terms of three parameters $\sigma_R, \sigma_I$ and $T_{corr}$. 
Given the baseline configuration of the observation, the uv-grid size chosen, and the integration time, we simulate the visibilities with different values of the parameters $\sigma_R, \sigma_I$ and $T_{corr}$ for each of the point source sky models for $8$ hours of total observation time. We also estimate the ideal visibilities expected from each sky model while no gain errors or measurement noise is present. The residual visibilities are calculated by subtracting the ideal visibilities from the simulated ones. \footnote{ Note that here ideal visibilities represent a known foreground sky model. In practice, such foreground models are derived from observations with their associated uncertainties. Hence, the results presented here can be considered as the best-case scenario when the foreground estimation is robust.} We then estimate the power spectra of the residual visibilities for all $128$ sky models. The mean of the power spectra from $128$ realizations provides an estimate of the bias in the power spectrum arising from residual gain errors. The variance of the power spectra from $128$ realizations gives an estimate of the excess variance arising due to residual gain errors. These are then compared with analytical calculations done with the same gain error models, baseline pair fractions of the observation, and measurement noise.

We show the relative contributions of different terms in the expression of $\sigma_P^2$ in Figure~\ref{fig:sigcomp}. For the set of parameters used here, the contribution from the gain error (Term 3) dominates over the system noise (Term 1). However, this behavior changes for different choices of the gain parameters and will be discussed shortly. Interestingly for the particular case shown here, the contribution in the uncertainty from different terms have similar baseline dependence. This indicates that the baseline dependence is mostly a result of the variation of total baseline pairs present in an annulus ($N_B$) as a function of baseline and depends less on the other baseline dependent factors.

Figure~\ref{fig:gcomp1} show the comparison of the analytical results with that of the simulation of bias and variance in the power spectrum estimator for different gain model parameters. The bias $\mathcal{B}_P$ is plotted with the black lines in the left panel, and the grey lines show the standard deviation $\sigma_P$ in the right panel as a function of the baseline. The thick curves are for analytical estimates and are denoted in the legend with `A'. As denoted as `S' in the legend, the thin curves show the corresponding results with simulations.
In top panel of the figure~\ref{fig:gcomp1}, we keep $\sigma_R = 1 \ \%$ and $T_{corr} = 16$ sec, same as the integration time for this observation. The solid and dashed lines show for $\sigma_I = 1.0\ \%$ and $5\ \%$ respectively. In the plot (a) we show the bias $\mathcal{B}_P$ and in (b) we show the standard deviation $\sigma_P$. We observe that $\mathcal{B}_P$ is significantly lower than $\sigma_P$ for $\sigma_I = 1.0\ \%$ and it is comparable in case of $\sigma_I = 5.0\ \%$. Increase in $\sigma_I$ also increases the bias. For these gain error parameters, the variance is dominated by the system noise and hence does not vary much with gain error models. A $\sigma_R = \sigma_I = 1\ \%$ corresponds to $1.0\ \%$ error in estimation of the amplitude and $0.6 ^{\circ}$ error in estimation of the phase of visibilities. Similarly for $\sigma_R = 1\ \%, \sigma_I = 5\ \%$, these numbers will corresponds to $1.0\ \%$ error in estimation of the amplitude and $3 ^{\circ}$ error in estimation of the phase.
 
In the bottom panel of figure~\ref{fig:gcomp1}, we show the variation of $\mathcal{B}_P$ in (c) and $\sigma_P$ in (d) for different values of $T_{corr}$ while $\sigma_R$ and $\sigma_I$ are kept at $1\ \%$ and $5\ \%$ respectively. We see that both $\mathcal{B}_P$ and $\sigma_P$ increases with increase in $T_{corr}$. For these gain model parameters change in $\mathcal{B}_P$ is more than $\sigma_P$ for different $T_{corr}$ and $\mathcal{B}_P$ exceeds the $\sigma_P$ for $T_{corr} = 64$ seconds. If we keep $\sigma_R$ and $\sigma_I$ at the same value, we shall refer to it as $\sigma_g = \sigma_R = \sigma_I$ henceforth.

Figure~\ref{fig:gcomp1} demonstrates that the variance estimates from simulation match with their analytical expression quite well. The analytical expression of the bias follows the estimates from simulation with a slight offset for some of the gain model parameters. However, these departures are relatively small. We conclude that the analytical expression for the bias and variance of the power spectrum we have presented here are in good agreement with the results from the simulation. This allows us to assess the effect of residual gain errors for different gain error models without referring to simulated observations, and hence reduces the computation time significantly. 

\section{Different effects of gain errors}
\label{sec:section4}

\begin{figure}
    \centering
    \includegraphics[width=0.5\textwidth]{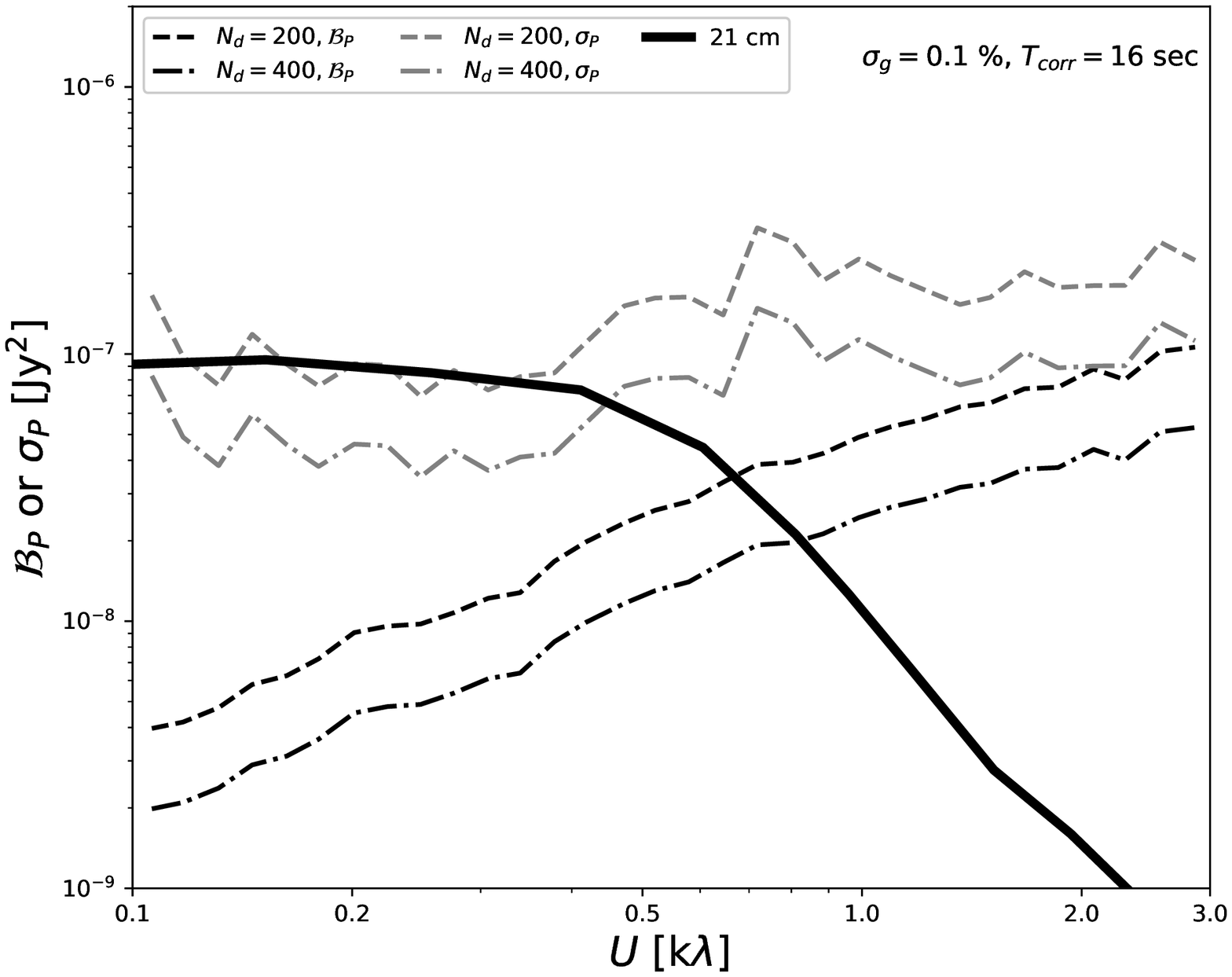}
    \caption{Variation of $\mathcal{B}_P$ (thin black) and $\sigma_P$ (thin grey) with baseline is shown for $100$ and $400$ days of observing time. Here, we consider a bandwidth of $16$ MHz, and a correlation time of $T_{corr} = 16$ seconds. Both the quantities $\sigma_R$ and $\sigma_I$ are set to $\sigma_R = \sigma_I = \sigma_g$. The thick black line corresponds to the expected EoR power spectrum \citep{2005MNRAS.356.1519B}. }
    \label{fig:time}
\end{figure}
In paper 1, we use a slightly different model for the gain error, where the normalized two-point correlation function of the residual gains is assumed to be unity at zero delay and falls of smoothly as a power-law at the larger delay. There we estimate the bias in the power spectrum using simulated observations. The gain error model discussed in this work is distinct from that of paper 1 in the sense that it uses an exponential function for the two-point correlation of the residual gains. However, they both assume the two-point correlation function of the residual gains to be unity at zero delay and zero at the large delay. In this work, we establish an analytical expression that can be used to estimate the effect of gain error in power spectrum for the gain error model discussed in section \ref{subsec:Gmodel}. In this section, we use the analytical expression for the bias and variance of the power spectrum to investigate the various effects of gain errors. It is important to note that the estimates of the bias and variance of the power spectrum presented in this work are based on a simplified model for gain error and by no means complete. We expect to have different effects, sometimes larger than what is discussed here, arising from various other sources like long time correlation, non-Gaussian effects and frequency correlation in gain errors, an inaccurate estimate of the sky model, foreground, etc. Hence, the numbers presented here should be taken only in the context and model of the gain errors discussed here.

Assuming that the gain errors are not correlated in frequency and across different days of observation, we first investigate the time required to detect the EoR 21-cm signal at $150$ MHz in the presence of residual gain errors. Here we have considered an observation bandwidth of $\Delta \nu = 16$ MHz with $\Delta \tau = 16$ sec of time integration per visibility. The bandwidth directly affects the system noise of the interferometer, whereas the integration time affects the system noise as well as the total number of baseline pairs in a given uv-grid where the power spectrum is estimated. For most of the analysis presented henceforth, we use the above values of $\Delta \nu$ and $\Delta \tau$. We shall discuss the effect of the integration time later. We observed in Figure~\ref{fig:gcomp1} earlier that the gain error increases with both $\sigma_R$ and $\sigma_I$ as well as $T_{corr}$. Here we choose a relatively moderate value for $\sigma_g = 0.1$ \% and correlation time $T_{corr} = 16 $ seconds. Using the analytical expression given in section \ref{subsec:AnalyticBP}. we calculate $\mathcal{B}_P$ and $\sigma_P$ for different number of days of observations $N_d = 200, 400$ and the result is shown in fig \ref{fig:time}. With the combinations of $\sigma_g$ and $T_{corr}$ discussed here, we observe that $\mathcal{B}_P$ is consistently less compared to $\sigma_P$ at all baselines. Hence, for the parameter values given here, detection of redshifted 21-cm signal would depend on the modified $\sigma_P$. The thick black curve in this figure shows the expected 21-cm power spectrum \citep[adapted from][]{2005MNRAS.356.1519B}. The 21-cm power spectrum at this observing frequency remains almost constant up to a baseline of $\sim 0.4$ k$\lambda$ and falls rapidly at higher baselines. If the gain error model presented here were the only source of power spectrum estimation error, we see that for observation of $200$ days $\sigma_P \sim P_{HI}$ for $U<0.4$ k$\lambda$. As $\sigma_P \sim 1/N_d$, we expect a $3-\sigma$ detection of the power spectrum for $600$ days of observation. Note that, as discussed earlier, in practice, there are expected to be more systematic effects other than what is discussed here that may require more stringent observational requirements. 
If excellent bandpass calibration is achieved, the effect of time-correlated gain error could become a dominant calibration error mechanism. Accuracy of the foreground model, efficacy of foreground subtraction procedure, unmitigated residual RFI, polarisation leakage, non-Gaussianity in noise, amongst other things that can also increase the uncertainty in power spectrum estimate.

\begin{figure}
    \centering
    \includegraphics[width=0.5\textwidth]{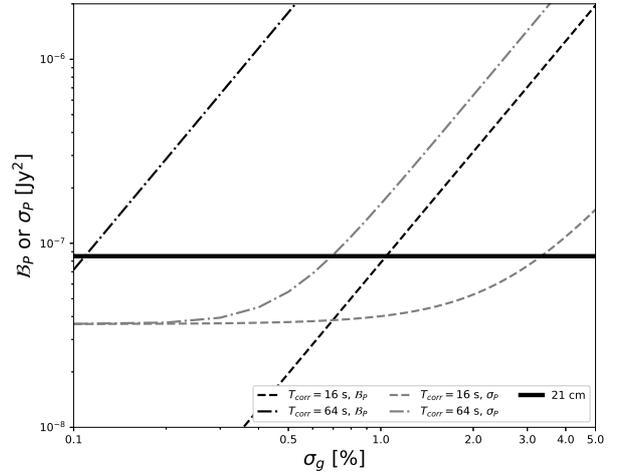}
    \caption{Variation of $\mathcal{B}_P$ (thin black) and $\sigma_P$ (thin grey) with $\sigma_R = \sigma_I = \sigma_g$ for correlation time $T_{corr} = 16, 64$ seconds. Here, we consider a bandwidth of $16$ MHz, and an observation time of $400$ days. The plots are made at a baseline of $300$ $\lambda$. The thick black line corresponds to the expected EoR power spectrum \citep{2005MNRAS.356.1519B}}
    \label{fig:sigG}
\end{figure}

For the compact source foreground model considered here, we observe that the baselines $>1$ K$\lambda$ are relatively more affected by the residual gain errors. This is partly because the $\mathcal{B}_P$ is systematically higher at longer baselines. However, the main reason may be that the amplitude of the expected redshifted 21-cm power spectrum reduces drastically beyond $0.4$ k$\lambda$. In the rest of the discussion, we investigate the effect of residual gain errors at the baseline of $0.3$ k$\lambda$. We also consider $\Delta \nu = 16$ MHz, $\Delta \tau = 16$ sec and a total of $400$ days of observation in the subsequent cases discussed here.

Figure~\ref{fig:sigG} show the variation of $\mathcal{B}_P$ (thin black lines) and $\sigma_P$ (thin grey lines) with $\sigma_g$ at a baseline of $0.3$ k$\lambda$. The thick horizontal black line corresponds to the expected EoR power spectrum at this baseline. The dashed curves show results with $T_{corr} = 16$ sec, whereas the dot-dashed curves show results with $T_{corr} = 64$ seconds. In eqn~\ref{eq:BiasA}, the term $\sigma_R^2+ \sigma_I^2 = 2 \sigma^2_g$ and hence 
it is expected that the bias increases with the value of $\sigma_g$ quadratically. This can be clearly seen in the figure. For small values of $\sigma_g$, the third term in the expression of $\sigma_P$ in  eqn~\ref{eq:VarA} dominates and it remains constant at well below the EoR power spectrum. We see that for $\sigma_g = 0.5$ \% or higher the effect of $\sigma_g$ is more pronounced. We also observe that at lower values of $\sigma_g$, $\sigma_P$ has a greater effect than the bias, as the gain error increases though, the power spectrum estimates become significantly biased. We observe that for cases with lower $T_{corr}$, i.e, for $T_{corr} = 16$ sec, contribution from $\chi(U)$ is rather small and both $\mathcal{B}_P$ and $\sigma_P$ are lower.

\begin{figure}
    \centering
    \includegraphics[width=0.5\textwidth]{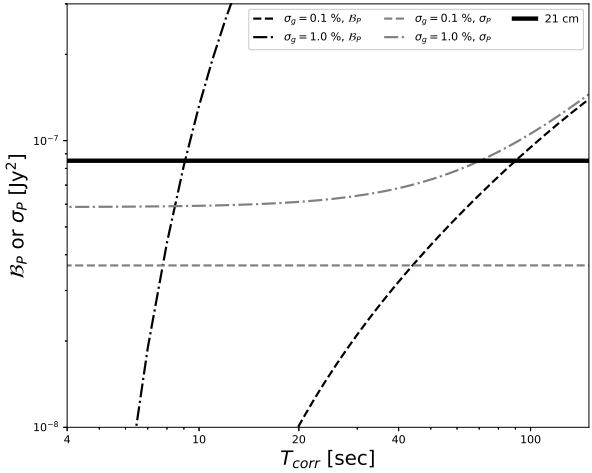}
    \caption{Variation of $\mathcal{B}_P$ (thin black) and $\sigma_P$ (thin grey) against  different correlation time with $\sigma_R = \sigma_I = \sigma_g = 0.1, 1.0$ \%. Here, we consider a bandwidth of $16$ MHz, and an observation time of $400$ days. The plots are made at a baseline of $300$ $\lambda$. The thick black line corresponds to the expected EoR power spectrum \citep{2005MNRAS.356.1519B}.}
    \label{fig:Tcorr}
\end{figure}

We investigate the effect of the correlation time $T_{corr}$ in Figure~\ref{fig:Tcorr}, where
we keep $\sigma_g$ fixed and vary $T_{corr}$ from $4$ to $128$ seconds. Note that, the integration time is fixed at $16$ seconds for all these cases. The effect of $T_{corr}$ comes through the function $\chi$ in both $\mathcal{B}_P$ and $\sigma_P$. The terms in the expression for $\mathcal{B}_P$ and $\sigma_P$ involving $\chi$ depends on the baseline pair fraction of type 1 and 3. Hence, for an interferometer baseline configuration with these baseline pair types relatively low, $T_{corr}$ would have lesser effect. We observe that for small values of $\sigma_g$, $\sigma_P$ is rather small, below the expected EoR signal and almost remains independent of $T_{corr}$. When $\sigma_g = 1$ \% is considered, $\sigma_P$ is still small at lower $T_{corr}$, but then increases drastically beyond $32$ seconds. The bias in the power spectra, on the other hand is rather large and well above the EoR signal for $T_{corr}< \Delta \tau$ with $\sigma_g = 1$ \%. For both the cases of $\sigma_g$, there is a transition point for $T_{corr}$ where $\mathcal{B}_P$ becomes dominant over $\sigma_P$.

\begin{figure}
    \centering
    \includegraphics[width=0.5\textwidth]{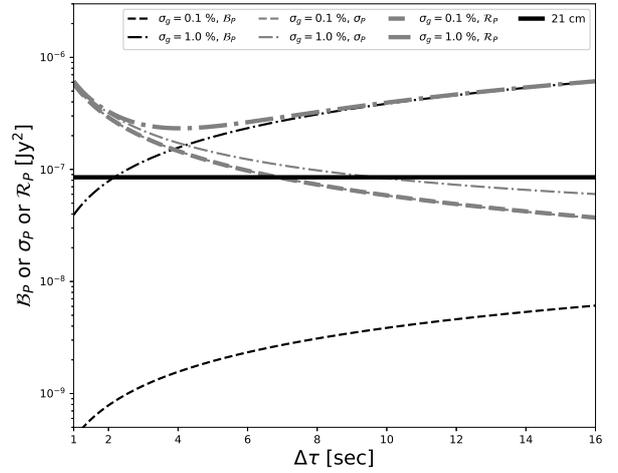}
    \caption{Variation of $\mathcal{B}_P$ (thin black) and $\sigma_P$ (thin grey) against different integration time $\Delta \tau$ with $\sigma_g = 0.1, 1.0$ \%. Here, we consider a bandwidth of $16$ MHz, and an observation time of $400$ days. The plots are made at a baseline of $300$ $\lambda$. The correlation time for the gain errors are kept same as $\Delta \tau$. The thick black line corresponds to the expected EoR power spectrum \citep{2005MNRAS.356.1519B}. The thick grey lines show the values of Risk $\mathcal{R}_P$ as defined in the eqn~\ref{eq:risk}}.
    \label{fig:Tint}
\end{figure}

In interferometric calibration, we estimate the gain as a function of time and try to reduce its effect from the observed data. Hence, the time correlation in the residual gain errors is not expected to be present at the time interval the gains are estimated. As each visibility measurement is integrated over the time $\Delta \tau$, the time variation can be estimated at best to a time scale of $\Delta \tau$. Hence, any time correlation in the residual gain errors can be at best reduced for $T_{corr} > \Delta \tau$. The maximum integration time for any interferometer is limited by the time smearing effect that arises from its largest baselines owing to integration time. For the GMRT, the maximum integration time that can be safely taken is $\Delta \tau = 16$ seconds. As one lowers $\Delta \tau$, there are two competing effects. On the one hand, the noise in each visibility increases (see eqn~\ref{eq:visnoise}). At the same time, the number of baseline pairs in a given uv-grid $N_B$ also increases. Hence, getting a reliable calibration solution for an arbitrary small integration time is difficult. This suggests that there can be an optimum integration time choice possible for a given observation and other parameters. We show the variation of $\mathcal{B}_P$ (thin black) and $\sigma_P$ (thin grey) with $\Delta \tau$ in the Figure~\ref{fig:Tint} for two values of $\sigma_g = 0.1, 1.0$ \%. We observe that the bias increases monotonically as integration time increases irrespective of the value of $\sigma_g$; though, for lower $\sigma_g$, the bias is negligible. The value of $\sigma_P$ is rather high for lower integration times and decreases with the increase of $\Delta \tau$. This is because at lower values of $\Delta \tau$
the contribution from the last term in the expression of $\sigma_P$ dominates through the foreground factor. The third term becomes important at larger $\Delta \tau$ through $\sigma_N/N_B$. We see that for $\sigma_g = 1$ \%, this eventually makes $\sigma_P$ fall below $\mathcal{B}_P$. We define the risk \footnote{Note that this is different from the usual definition of risk. Here we use this to keep it of the same dimension of $\mathcal{B}_P$ (thin black) and $\sigma_P$.} as 
\begin{equation}
    \mathcal{R}_P^2 = \mathcal{B}_P^2 + \sigma_P^2,
    \label{eq:risk}
\end{equation}

The risk for the two cases shown here is plotted with thick grey lines. As the bias for the case with $\sigma_g = 0.1$ \% is significantly low, the risk mostly follow its $\sigma_P$. On the other hand, for the case with $\sigma_g = 1.0$ \%, we see that risk is at its lowest for an integration time of $4$ seconds. This demonstrates that for such high dynamic range observations, one needs to assess the gain properties of the telescope and can optimally choose the integration time for observation. Note that it is always possible to use the lowest integration time; however, as this exercise demonstrates, one needs to integrate the visibilities to an optimal time prior to using visibility correlation.

The regular interferometric calibration uses observation of calibrator sources to assess the large time scale variation of the gain. However, one uses the self-calibration procedure to mitigate the time dependence of the gain at a few tens of seconds time scale. Here one may reduce the time interval at which the gain solution is attempted and get a gain solution at smaller time scales. This is expected to reduce the correlation time of the residual gain errors. However, estimating gain solutions at higher time resolution reduces the number of independent measurements used to estimate the solution increasing the calibration uncertainty. In addition to the sky model, in self-calibration, the amplitude and phase closure properties are used \citep{1986isra.book.....T} to constrain the gain solutions of antenna. Hence, given the telescope SEFD, the integration time of observation $\Delta \tau$, the amplitude of the signal used for calibration $A_0$, the time cadence at which the gains are estimated $T_{sol}$ and the number of antenna $N$ available to impose the closure properties in phase and amplitude, in the ideal case the variance of the real and imaginary parts of the residual gain errors differ and can be written as $\sigma_I = \sigma_g, \sigma_R = \eta\ \sigma_g$, where
\begin{equation}
\sigma_g^2 = \frac{2 \sigma_N^2 /A_0^2}{(N-1) (N-2) T_{sol}/\Delta \tau},
\label{eq:selfcal}
\end{equation} 
and $\eta^2 = 3/(N-3)$. In the best possible case, the gain solutions can be obtained at the time interval of the integration, and any time correlation in the residual gain would have a lower correlation time, that is $T_{corr} = T_{sol} = \Delta \tau$. Considering a value of  $A_0^2$ equal to the amplitude of the foreground power spectrum, the typical value of $\sigma_g \sim 0.5$ \% for $16$ seconds of integration time and $62.5$ kHz channel width with the uGMRT. As $\eta$ is small, the main contribution to the bias and variance of the power spectrum arises from the imaginary part of the gain errors.
\begin{figure}
    \centering
    \includegraphics[width=0.5\textwidth]{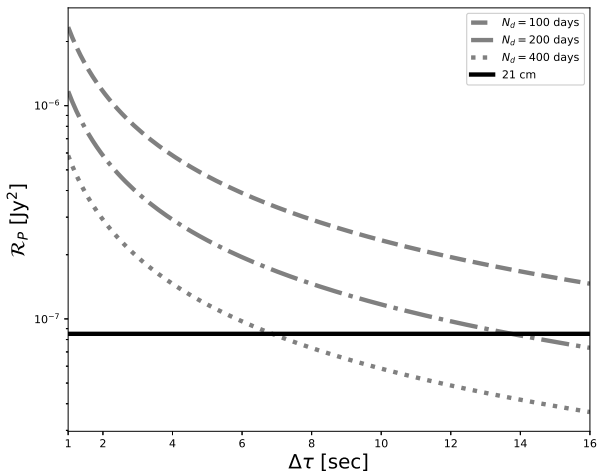}
    \caption{Variation of $\mathcal{R}_P$ against different integration time $\Delta \tau$ for the best possible case with the uGMRT at $150$ MHz. Here, we consider a bandwidth of $16$ MHz, and an observation times of $100, 200$ and $400$ days. The plots are made at a baseline of $300$ $\lambda$. The thick black line corresponds to the expected EoR power spectrum \citep{2005MNRAS.356.1519B}. }
    \label{fig:best}
\end{figure}

Figure~\ref{fig:best} show the estimates of the risk for three different total observation days with $8$ hours of observation time per day for the best calibration possible with the uGMRT. We see here that the risk monotonically decreases, and the best choice is the maximum integration time allowed in the uGMRT that avoids time smearing. As the expression in eqn~\ref{eq:selfcal} suggests, the risk is expected to be less for an interferometer with more antenna and hence more baselines available for estimation of the gains. For two interferometers with the same sensitivity, the one with a larger number of antennas would have a lesser effect from the residual gain errors.

\section{Discussion and Conclusion}
\label{sec:section5}

The presence of strong foreground at the observing frequencies of redshifted 21-cm radiation makes detection of the cosmological \HI signal more challenging. 
Calibration error couples with the strong foreground signal and limits the possibility of the 21-cm signal detection. In this work, we addressed the issue of estimating the redshifted 21-cm power spectrum in the presence of strong foreground and residual gain errors from calibration. Using a model for the gain errors in radio interferometric observation, we derive analytic expressions for bias and variance of the power spectrum by propagating the uncertainties in visibilities from thermal noise and gain errors. This analytical expression is then tested against simulated observations from the GMRT, where we see a fairly good agreement between the two. As in paper 1, we have already demonstrated through simulated observation that the bias in the power spectrum arising from time correlations in the gain errors alone is important to estimate to assess the statistical significance of the 21-cm power spectrum measurements. However, such a simulated observation, particularly with the advent of new telescopes with a significantly larger number of antennae and angular resolution, is computationally challenging. Furthermore, assessing the statistical measures through simulation requires running multiple realizations of the various components with random uncertainties, which increases the computational load. The presented analytical formula here provides a much more efficient way of estimating the bias and variance of the power spectrum. Once established by comparison with the simulated observations, we use the analytical expression of the bias and variance to assess the effect of different components of the gain error model and see their relative effects on the bias and variance. The later quantities depend on the baseline configuration of the telescope. The comparative studies done here assume the uGMRT baseline properties for reference. Similar analysis can be carried out for any array configuration and telescope properties as the need may arise.

The gain error model considered here assumes that the residual gain error is Gaussian random with given variance for its real and imaginary parts and an auto-correlation function with a finite correlation time. We find that the bias in the power spectrum strongly depends on both the variance of the gain errors and the correlation time, whereas the variance in the power spectrum is more sensitive to the variance in the gain errors. We find that for the uGMRT, even with variance in the gain error as low as $1$ \%, the bias in power spectrum exceeds $\sigma_P$ for a reasonably low correlation time of the gain error. As an example, we observe that with a $16$ MHz of bandwidth and $3200$ hours of observations, for $T_{corr}>10$ seconds, the bias in the power spectrum is significantly higher than its $\sigma_P$. This suggests that it is important to assess the bias in the 21-cm power spectrum estimation in the presence of strong foreground, failing which a biased estimate can confuse the scientific interpretation of the signal. The properties of noise in individual visibilities, as well as the correlation time of gain errors, are expected to be affected by the integration time for each visibility in an observation. We find that, for a moderately low variance in gain errors, the risk in such an observation can be minimized by choosing a moderate value for the integration time, hence increasing the possibility of an unbiased detection.

This work is the second in a series of works aimed at understanding the effect of residual gain errors in different power spectrum estimations in the presence of strong foreground and exploring potential mitigation techniques. In this work, we have not investigated the different possibilities for the presence of residual gain errors and chosen the values of $\sigma_R$ and $\sigma_I$ same as $\sigma_g$ for most of the discussions. In general, the standard deviation of the real and imaginary parts can be different. We observe here that it is essential to assess the time dependence of the gain accurately, as its inaccurate estimation leads to the time-correlated residual gain errors. 
In this work, we do not consider the effect of frequency correlation in gain error, and all our estimations are done for correlating visibilities in the same frequency channel. Furthermore, this work also uses the foreground subtraction technique, where we expect to have accurate knowledge of the foreground emissions \citep{2008MNRAS.389.1319J, 2012MNRAS.426.3295G}. An alternative method, more regularly exercised in literature, is foreground avoidance. It has been established that the foregrounds to the redshifted 21-cm emissions remain correlated across relatively larger bandwidth \citep{1998ApJ...505..473P, 2005ApJ...625..575S, 2008MNRAS.385.2166A, 2008MNRAS.389.1319J, 2019MNRAS.490..243C}, whereas the \HI signal decorrelates faster \citep{2003JApA...24...23B, 2005MNRAS.356.1519B}. As a result, when the power spectrum is observed as a function of $(k_{\parallel}, k_{\perp})$, the foreground emission remains concentrated near the low $k_{\parallel}$, inside the `wedge' \citep{2010ApJ...724..526D, 2012ApJ...752..137M, 2012ApJ...745..176V}. Note that the smaller frequency separation in multi-frequency angular power spectrum contributes to larger $k_{\parallel}$ modes of the power spectrum. Hence, the effect of frequency-independent residual gain we see here at zero frequency separation may contribute to bias in the power spectrum beyond the wedge. Moreover, the antenna-based gains are functions of both time and frequency; the residual gain is expected to have correlated frequency dependence.
Such frequency-correlated calibration errors couple the foreground power beyond the foreground wedge into the EoR window region of the 2-D power spectrum space \citep{2016MNRAS.461.3135B, 2017MNRAS.470.1849E, 2019ApJ...875...70B, 2021MNRAS.501.3378P}.
At present, we are working towards expanding the formalism presented in this paper to estimate bias and variance in power spectrum estimate when visibility correlation in different frequencies is considered.
Here, we also consider that the gain errors arising from the different antennae are uncorrelated. Though this is a fairly good assumption for the gain arising from electronics in the antenna system itself, the ionospheric effects may introduce correlated gains across the antenna. Furthermore, as the calibration procedure uses baseline dependent gains to solve for the antenna dependent gains, calibration errors can lead to correlated residual gain errors across the antenna. Moreover, asymmetry in the telescope aperture, mechanical fatigue of telescope structure, etc., can lead to parts of the gain errors correlated across different antennae and even across different days of observations. We are investigating these effects, and the result will be presented in future work.
Though the demonstrations here are done with the uGMRT as a model for the interferometer, a similar analysis can be carried out for any telescope of concern, and a prior assessment of the effect of the gain errors can be made using the analytical expression presented here with minimum computation cost. Furthermore, this work emphasizes the importance of estimating and establishing the gain statistics for a given interferometer. Though we use a simple model for the residual gain error here, the calculations that lead to the analytical expression can be readily expanded for a more complicated gain error model. We believe this work provides significant direction in understanding and planning observations to detect redshifted 21-cm power spectrum.

\section*{ACKNOWLEDGMENT}
JK would like to acknowledge the University Grant Commission (UGC), Government of India, for providing financial support through the Senior Research Fellowship. PD acknowledges discussion with Wasim Raja about various aspects of this work. SC would like to thank Philip Bull for useful discussions. The support and the resources provided by ‘PARAM Shivay Facility’ under the National Supercomputing Mission, Government of India at the Indian Institute of Technology, Varanasi, are gratefully acknowledged. The authors thank the anonymous referee for suggestions that have improved the presentation of the paper significantly.

\section*{DATA AVAILABILITY}
No new data were generated or analyzed in support of this research.

\bibliographystyle{mnras}
\bibliography{references.bib}

\end{document}